# Searches for physics beyond the standard model at the LHC

Jory Sonneveld for the CMS collaboration

*Institute for Experimental Particle Physics, University of Hamburg, Luruper Chaussee 149, 22761 Hamburg, Germany*

At the Large Hadron Collider (LHC) at the European Organization for Nuclear Research (CERN), protons and heavy ions are accelerated to velocities close to the speed of light and collided in order to study particle interactions and give us an insight to the fundamental laws of nature. The energy and intensity of the particle beams at the LHC are unprecedented, and a tremendous amount of data is collected by three experiments on the circular ring of the LHC that are specialized in proton-proton collisions. The data confirm the most successful theory of particle physics to date known as the standard model of particle physics to very good precision, including the long expected and recently discovered Higgs boson. The standard model cannot, however, accommodate experimentally observed phenomena like gravity, neutrino masses, and dark matter. The theory can also be theoretically unsatisfying as a result of parameters that go unexplained, such as the relatively low value of the Higgs mass despite its large quantum corrections, implying a lack of understanding. For this reason, in addition to precision measurements of standard model observables, experiments search for new physics beyond the standard model that could explain some of the shortcomings of the standard model. A selection of results for searches for new physics beyond the standard model using data recorded by three experiments on the LHC are presented in this talk.

## 1. Introduction

On average about ten thousand particles from cosmic rays pass through each square meter every second. The HESS experiment in Namibia can detect such cosmic rays from our galaxy where protons can be accelerated to energies of the order of peta-electronvolts[1] [1]. Experiments in particle and astroparticle physics are used to answer questions about what these particles are made of, as well as what we and the universe are made of. Instead of waiting for particles from outer space, the Large Hadron Collider [2] at CERN in Geneva collides its own bunches of $10^{11}$ protons at 6.5 tera-electronvolt (TeV) per beam up to 1 billion times a second. This is the largest and most powerful collider in the world. Seven detectors on the 27-kilometer ring are used to collect data from the 4 collision points on the LHC. Results from analyses of data collected by three of these are discussed. Two multi-purpose detectors, A toroidal LHC ApparatuS (ATLAS) [3] and a Compact Muon Solenoid (CMS) [4], located on opposite sides of the LHC ring, are both competitive and complementary in measurements of the established and new models of physics. CMS weighs 14000 tons, about twice as much as ATLAS and 1.5 times the weight of the Eiffel tower, and is with its 15m diameter, 25m width and 46m length only about half the size of ATLAS, which has a very good jet resolution. CMS has the most powerful solenoid magnet ever made and excellent particle momentum resolution in its silicon tracker. LHCb [5] is a forward-arm spectrometer specialized in studies of the bottom and charm quarks. It has a very accurate tracking system starting from only 8.2mm from the beam pipe, and is very forward[2] ($2 < |\eta| < 5$) compared to CMS and ATLAS ($|\eta| < 2.4$) making it sensitive to the region where b-jets are predominantly produced (see Fig. 1-3).

*Figure 1: A b-jet event in LHCb; figure from* https://pbs.twimg.com/media/Db5WuWEXkAAxGIY.jpg

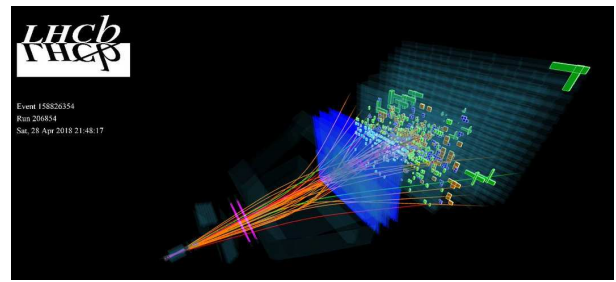

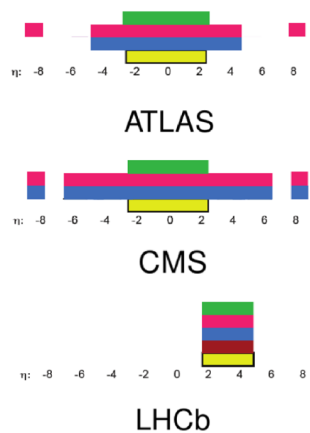

Figure 2: The forward position of LHCb. Yellow represents tracking detectors, blue electromagnetic calorimeters, pink hadronic calorimeters, brown detectors for hadron particle identification, and green muon systems. Figure from https://inspirehep.net/record/1398587.

---

[1] In particle physics, energy, momentum and mass are often expressed in electronvolt. One electronvolt is the amount of energy lost or gained by a particle of a single electron charge (1*e*) moving across an electric potential difference of one volt (1 V) and is approximately 1 eV ≈ 1.602 × 10$^{-19}$ J (joule) in SI units.

[2] At the LHC, pseudorapidity η is used to describe the angle of a particle with respect to the beam axis, and is defined as $\eta = -\ln(\tan(\theta/2))$, where θ is the polar angle in the CMS coordinate system.

The LHC and its experiments are designed to answer some of the fundamental questions of our universe such as what we are made of and how particles obtain mass. This and all known fundamental particles to date, as well as the electromagnetic, weak nuclear, and strong nuclear forces are very accurately described by the so-called standard model of particle physics [6–14]. However, it does not describe gravity or so-called dark matter, provide neutrino masses, or explain the matter-antimatter asymmetry in the universe. In addition, it can be theoretically unsatisfactory as the standard model, for instance, has huge corrections of the order of $(10^{19}$ GeV$)^2$ to the Higgs mass squared that we know should sum up together to be only $(125$ GeV$)^2$ when assumed to be correct to the Planck scale. A number of models of physics beyond the standard model (BSM) have been
proposed to answer some or all of the open questions in particle physics.

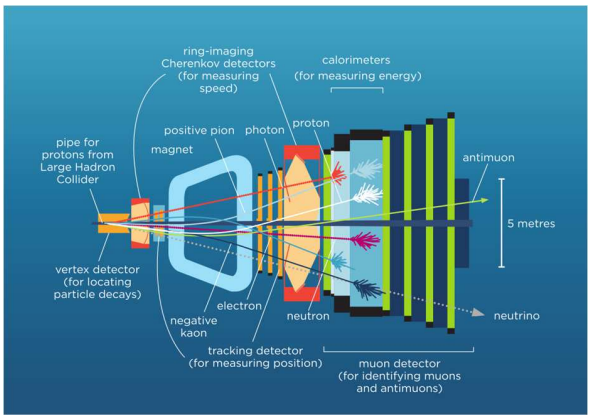

*Figure 3: The different subdetectors of LHCb with characteristic signatures of different particles; figure from http://antimattermatters.s3-website-eu-west-1.amazonaws.com/images/antimatter_web10.png.*

**2. Searches for physics beyond the standard model**

One can search for physics beyond the standard model with different techniques, for example through resonance searches (also known as bump hunting), searching for processes with associated large missing energy, looking for deviations in standard model observables, or studying angular distributions of final states. Missing transverse momentum, where, for example one jet is detected in one direction of an angle φ perpendicular to the beam pipe and obviously energy in the other direction is missing (see Figure 4), could be a sign of dark matter. It could, on the other hand, also be a detector effect or mismeasurement.

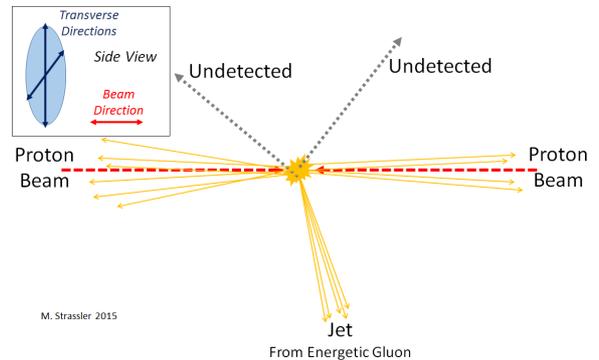

Figure 4: A proton-proton collision resulting in one jet and missing energy; figure from https://profmattstrassler.files.wordpress.com/2015/04/pp2nunuj_side.png.

**3. Resonances**

Examples of resonance searches are CMS-PAS-EXO-18-006 [15] where an invariant mass is computed from final state leptons for 41 fb$^{-1}$ collected[3] in 2017 (see Fig. 5). and ATLAS-CONF-2018-016 [16] where an invariant mass from an all-jet final-state is studied in ATLAS W- and Z- tagged events in 79.8 fb$^{-1}$ of data collected at the ATLAS detector in 2017 and 2018 (see Fig. 6). Results of both searches are compatible with the standard model in which case so-called exclusion (upper or lower) limits are computed on the model parameters using probabilities, see Fig. 7-8. Expected limits show what one would observe if the data behaved like the predicted background, and observed limits are computed with actual data. In this process one needs to be careful to avoid false positives, in which case one excludes the background (standard model) where it is actually correct, and false negatives, in which case one excludes the new hypothesis (BSM model) where it is actually correct. Backgrounds are derived either directly from data or from simulation, the latter of which demonstrates the importance of precise event generators and detector simulations.

---

3   At particle colliders, the cross section, or probability of a certain collision, is given per unit area and is typically expressed in barn. One barn is 10$^{-24}$ cm$^2$.

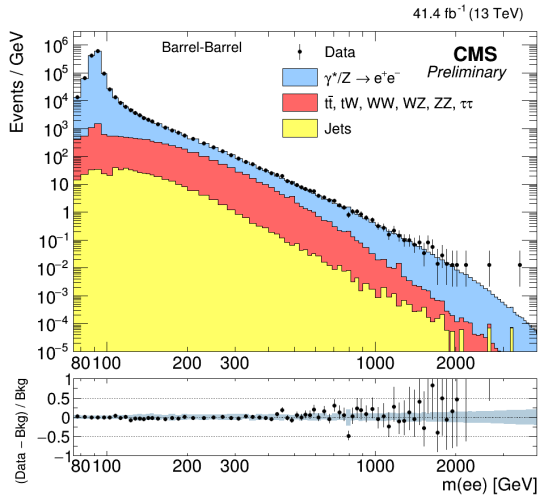

Figure 5: Dielectron invariant mass in ee events observed at the CMS detector; figure from [15].

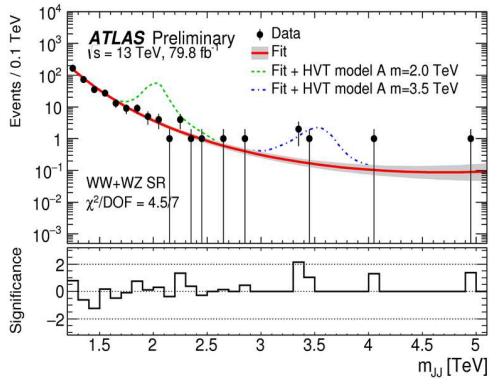

Figure 6: Dijet invariant mass in WZ- and ZZ-tagged events observed at the ATLAS detector; figure from [16].

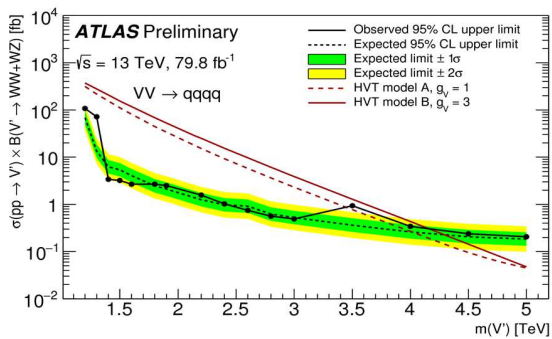

Figure 7: Upper limit on cross section times branching ratio for different masses of an extra-dimensional graviton in the WW- and ZZ-tagged channel; figure from [16]. Dashed represent expected limits, solid lines

observed limits. Expected cross sections from theory are also shown.

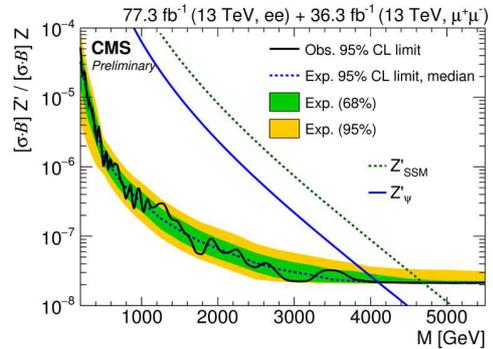

Figure 8: Upper limit on the production cross section times branching ratio of a spin-1 resonance relative to the Z-boson cross section times branching ratio for different resonance masses in the dielectron and dimuon channel; figure from [15]. predictions for two theories are overlaid (blue and dotted lines).

### 4. Deviations from standard model observables

Decay rates and cross sections of B mesons are determined by coupling strengths of interactions and are very sensitive to corrections from new particles that occur in models of new physics. Processes including B mesons often manifest themselves through a displaced vertex in the forward region as a result of the longer lifetime of the B meson. LHCb can measure such processes very well in its forward detector (see Fig. 9) and has recently published new-physics constraining results on the decay rate of the $B_s$ meson (which consists of one strange and one antibottom quark) to two muons using 4.4 fb$^{-1}$ of proton-proton collision data [17]. Deviations of these decay rates from standard model predictions could, for example, imply the existence of new Higgs bosons (see Fig. 10).

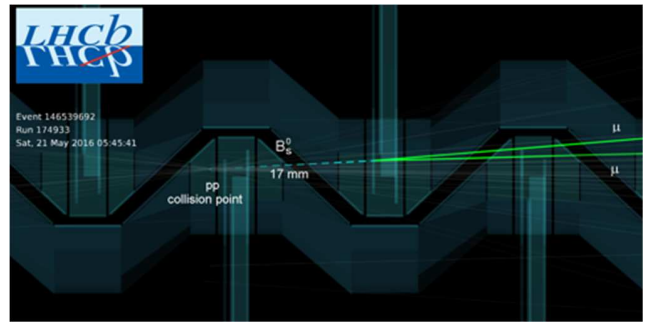

Figure 9: A displaced vertex from a $B_s \to \mu^+\mu^-$ decay in the LHCb detector; figure from https://lhcb-public.web.cern.ch/lhcbpublic/Images2017/BsMuMuVertex_s.png.

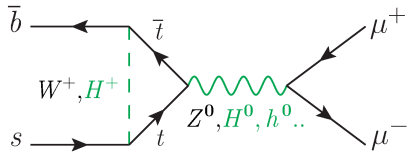

Figure 10: Feynman diagram of a $B_s \rightarrow \mu^+\mu^-$ process with standard model particles (black) and possible new Higgs particles (green). Figure from https://www.hep.physik.uni-siegen.de/atlas/pics/feynman_dig_Bs-MuMu.png

Results for the selected $\mu^+\mu^-$ pairs mass distribution and $B_s$, $B^0 \rightarrow \mu^+\mu^-$ are compatible with standard model predictions, as shown in Fig. 11-12. For this reason, these results are very constraining for the parameter space of models of physics beyond the standard model, as corrections to the $B_s \rightarrow \mu^+\mu^-$ branching fraction cannot be too large. The branching fraction of $B_s \rightarrow \mu^+\mu^-$ was measured to 7.8 standard deviations to be $(3.0 \pm 6^{+0.3}_{-0.2}) \times 10^{-9}$ assuming an amplitude ratio of 1 (as predicted by the standard model) of the $B_s$ and anti-$B_s$ meson decays.

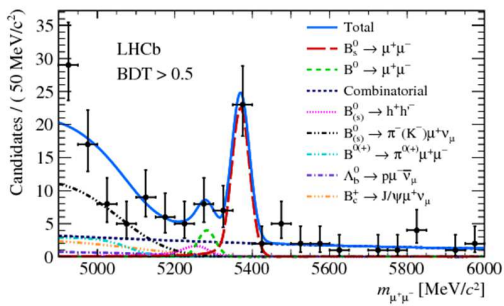

Figure 11: Mass distribution of $B_s \rightarrow \mu^+\mu^-$ muon candidates selected with a boosted decision tree algorithm; figure from [17].

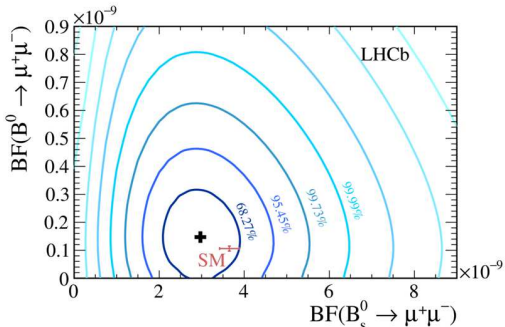

Figure 12: the measured branching fractions of $B_s \rightarrow \mu^+\mu^-$ and $B^0 \rightarrow \mu^+\mu^-$ as compared with the standard model expectation. Figure from https://lhcbproject.web.cern.ch/lhcbproject/Publications/LHCbProjectPublic/LHCb-PAPER-2017-001.html.

## 5. Tests of lepton universality

It has long been assumed that lepton universality, or the equality of interactions (couplings) of all leptons (all flavors) to gauge bosons, holds. LHCb results[4], however, have shown a deviation from the standard model prediction for a B meson decaying to a D* and a lepton and neutrino (see Fig. 13) in terms of the observable R(D*) as defined in Eq. 1. LHCb measured R(D*) to be 0.285 ± 0.019 ± 0.029 which is 2.1σ above the standard model prediction which is 0.252 ± 0.003. Together with several other measurements from experiments specialized in bottom quark mesons[5] the observed R(D*) is about four standard deviations away from the Standard Model.

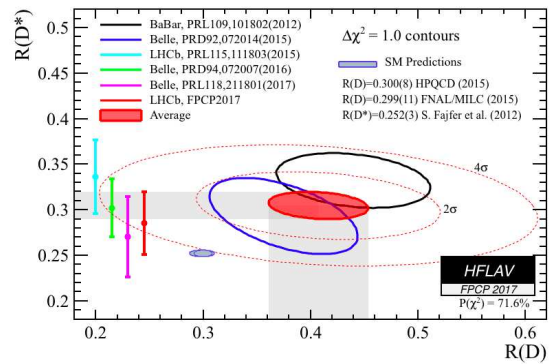

Figure 13: Worldwide results for measurements of the variable R(D*) as compared to the prediction from the standard model, where a deviation of about 4σ is observed. Note that some collaborations have also measured R(D), but for LHCb and Belle results one can imagine the measurements as a horizontal band as this measurement was independent of R(D). Figure from https://www.slac.stanford.edu/xorg/hflav/semi/fpcp17/rdtaunu/rdrds_fpcp2017.png.

$R(D^*) = (B^0 \rightarrow D^*\tau\nu_\tau) / (B^0 \rightarrow D^*\mu\nu_\mu)$ = (standard model) 0.252 ± 0.003. (Eq. 1)

These very exciting results show the most significant deviations from standard model predictions at the Large Hadron Collider to date.

---

4  See https://indico.cern.ch/event/586719/contributions/2531261/
5  See https://hflav-eos.web.cern.ch/hflav-eos/semi/fpcp17/RDRDs.html

## 6. Searches for dark matter at the LHC

Many galaxies would not hold together and rotate, or would not have formed the way they do if there were not a large amount of matter that we cannot detect. This and other evidence for such matter gives rise to the hypothetical 'dark matter' that makes up 85% of the matter of our universe. At particle colliders, the only way to look for dark matter is to look for processes where it is produced, as only non-dark matter particles are at hand to collide with, unlike in direct and indirect detection experiments that are designed to detect collisions including dark matter particles (see Fig. 14). Not all collisions can be stored continuously at ATLAS and CMS, and one way of deciding to read out event information (or so-called 'triggering') when 'nothing' is produced is to look for jets emitted through initial state radiation in events where then subsequently dark matter could be produced (see Fig. 14).

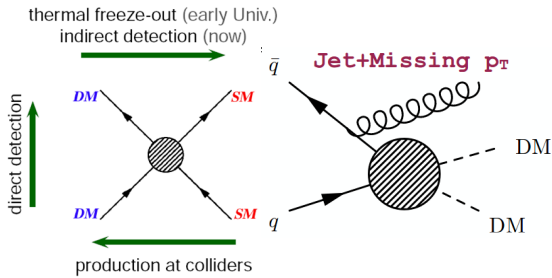

Figure 14: Different ways to search for dark matter (left, from https://www.mpi-hd.mpg.de/lin/images/research_theory5.png . At colliders one searches for production of dark matter particles; to be able to select events with such a process one relies on initial state radiation with for example a jet (right, figure from https://indico.cern.ch/event/528094/contributions/2172833).

One such monojet search by ATLAS [18] assumes a dark matter mediator (a particle that 'mediates' between standard model particles and particles from the dark matter or 'hidden' sector) and looks for an initial state radiation jet with a transverse momentum $p_T > 250$ GeV and an associated missing transverse momentum of $E_T > 250$ GeV. The search is performed using 36.1 fb$^{-1}$ of 13 TeV data collected with the ATLAS detector in 2015 and 2016. Their results are compatible with standard model predictions (see Fig. 15-16).

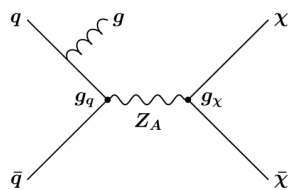

Figure 15: A search for dark matter with a Z mediator using, for example, initial state radiation from gluons. Figure from [18].

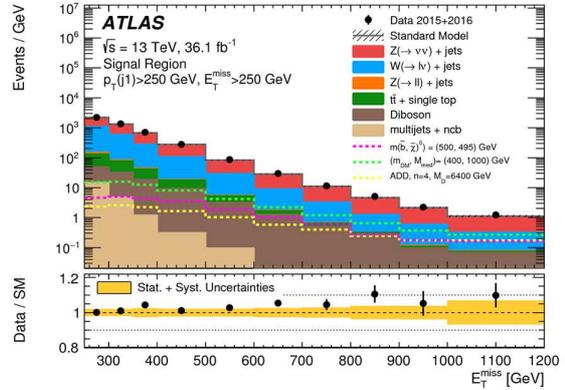

Figure 16: No deviation from standard model predictions was seen in the missing transverse energy distribution in this monojet search for dark matter in data collected by the ATLAS detector. Figure from [18].

## 7. Longlived particles

Unconventional signatures such as displaced vertices and heavy stable charged particles are also explored in BSM searches at the LHC (see Fig. 17). One such search in CMS recently presented at LHCP 2018 in Bologna looks for disappearing tracks [19]. No hint of such longlived particles has been observed in 38.4 fb$^{-1}$ of 13 TeV data collected at the CMS detector, but the searches proved to be sensitive to a large range of lifetimes and masses in a certain model of new physics (see Fig. 18).

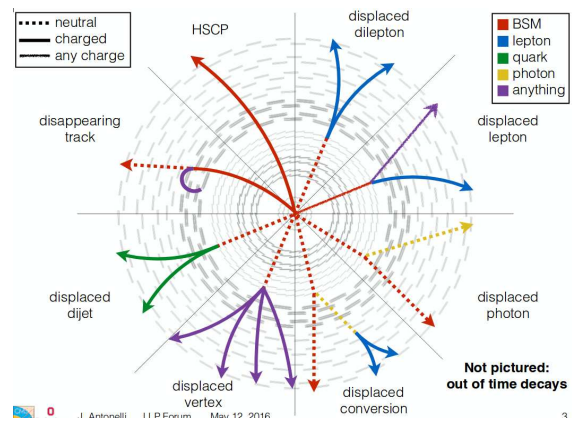

Figure 17: Possible signatures of longlived particles in the CMS and ATLAS detectors; figure from https://indico.cern.ch/event/517268/contributions/2041293.

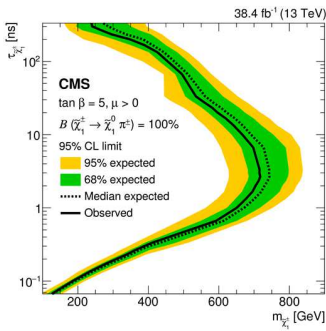

Figure 18: Limits on a BSM longlived particle called a chargino in a specific model of new physics from CMS which show the sensitivity of this search in this model's parameter space. Figure from [19].

## 8. BSM Higgs particles

The newly discovered standard model Higgs boson is studied in detail at the LHC. However, at the same time, particle physicists are looking for new BSM Higgs particles. One such CMS search looked for new Higgs bosons with four bottom quarks in the final state [20] in 35.7 fb$^{-1}$ of data collected at the 13 TeV LHC in 2016 (see Fig. 19). Another ATLAS search looked for production of H and Z bosons with the H decaying to two new Higgs particles denoted *a*, two leptons and four bottom quarks in the final state [21] in 36.1 fb$^{-1}$ of data collected in 2015 and 2016 at the 13 TeV LHC, and was able to interpret the results not only for such *a*'s decaying promptly but also for longer lived *a*'s. Neither ATLAS nor CMS observed a deviation from the standard model prediction (see Fig. 20).

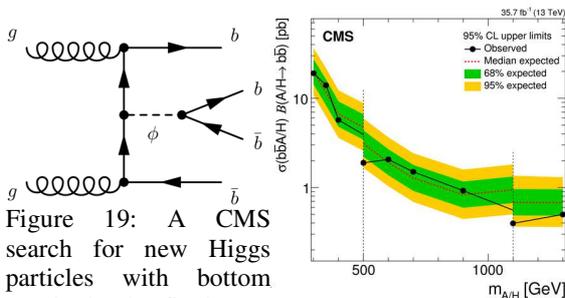

Figure 19: A CMS search for new Higgs particles with bottom quarks in the final state (left) found no deviation from standard model predictions and summarized its results in terms of the BSM Higgs mass and its cross section times branching ratio (right). The steps result from different selection criteria used in different mass subranges. Figures from [20].

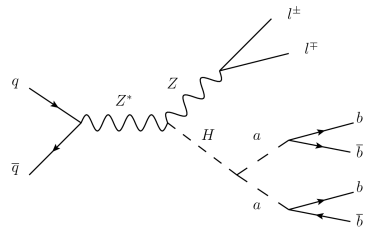

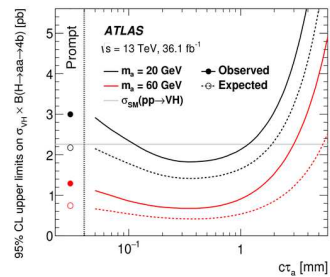

Figure 20: An ATLAS search for new Higgs particles with bottom quarks in the final state (top) found no deviation from standard model predictions and summarized its results in terms of the lifetime and cross section times branching ratio of so-called a bosons of mass 20 and 60 GeV (bottom). Figures from [21].

## 9. Conclusions

Searches for physics beyond the standard model of particle physics at the various experiments at the LHC aim to answer fundamental questions about our universe. Data taking at the LHCb, ATLAS and CMS experiments has been successful and analysis of those data gives an indication that the standard model is not blind to lepton flavor. There is no clear sign of new models of physics, but it is expected that many more results will follow as we gain more statistics in this and the future runs of the LHC.